\documentclass[prd,twocolumn,floatfix,amsmath,nofootinbib,amssymb,floatfix]{revtex4}
\usepackage{graphicx,color,dcolumn,booktabs,bm,multirow}
\usepackage{longtable,lscape}
\usepackage{txfonts}
\usepackage{overpic}
\usepackage{amssymb}
\usepackage{array}
\usepackage{indentfirst}
\usepackage{feynmf}   
\usepackage{slashed}  
\usepackage{cases}
\usepackage{color}
\usepackage{multirow}
\usepackage{epstopdf}
\usepackage{graphicx,color,dcolumn,booktabs,bm}
\usepackage[colorlinks,
citecolor=blue,
anchorcolor=red,
menucolor=red,
linkcolor=red,
filecolor=red,
runcolor=red,
urlcolor=blue,
frenchlinks=red]{hyperref}

\def\zcp{Z_c(4020)}
\def\zc{Z_c(3900)}
\def\psia{\psi(3770)}
\def\psib{\psi_2(3823)}
\def\psic{\psi_3(3842)}

\begin{document}
	
	\title{Pionic transitions from $Z_c(4020)$ to $D$ wave charmonia}
	\author{Xiao-Yu Qi$^{1}$}
	\author{Qi Wu$^{2}$}
	\author{Dian-Yong Chen$^{1,3}$\footnote{Corresponding author}}\email{chendy@seu.edu.cn}
	\affiliation{$^1$ School of Physics, Southeast University, Nanjing 210094, People's Republic of China}
	\affiliation{$^2$ Institute of Particle and Nuclear Physics, Henan Normal University, Xinxiang 453007, China}
	\affiliation{$^3$ Lanzhou Center for Theoretical Physics, Lanzhou University, Lanzhou 730000, China}
	\date{\today}

	\begin{abstract}
In the present work, we investigate the charmed meson loops contributions to the pionic transitions from $\zcp^+$ to the $D$ wave triplets charmonia by using an effective Lagrangian approach. Our estimations indicate that the predicted branching fraction of $\zcp^+ \to \pi^+ \psi(1^3D_J) , \ J=(1,2,3)$ are much smaller than the one of $\zcp^+ \to \pi^+ h_c $. Thus, searching $\zcp^\pm$ in the $\pi^\pm \psi(1^3D_J) $ invariant mass distributions is impossible. Thus, the observed peak structures at 4.04 and 4.13 GeV in the $\pi^\pm \psi(3770)$ invariant mass distributions should not come from the contributions of $\zcp^\pm$, and further precise experimental measurements of the $e^+ e^- \to \pi^+ \pi^- \psi(3770)$ process are needed to decode the nature of these two peak structures.
	\end{abstract}
	
	\maketitle

	\section{introduction}
	
	As the first confirmed charged charmonium-like state, $Z_c^\pm (3900)$, was firstly observed by the BESIII~\cite{BESIII:2013ris} and Belle~\cite{Belle:2013yex} Collaborations in the  $\pi^\pm J/\psi$ invariant mass spectrum of the process $e^+ e^- \to \pi^+ \pi^- J/\psi$ at $\sqrt{s}=4.260$ GeV one decade ago. Later on, the authors of Ref.~\cite{Xiao:2013iha} further confirmed the existence of $Z_c^\pm(3900)$ in the $\pi^{\pm }J/\psi$ invariant mass spectrum of the same process and also reported the first evidence of $Z_c^0(3900)$ in the $\pi^0 J/\psi$ invariant mass spectrum of the process $e^+ e^- \to \pi^0 \pi^0 J/\psi$ by using the data taken with the CLEO-c detector at $\sqrt{s}$=4.170 GeV. Later, the neutral partner, $\zc^0$ was observed in the process  $e^+ e^- \to \pi^0 \pi^0 J/\psi$ by the BESIII Collaboration~\cite{BESIII:2015cld, BESIII:2020oph}, and then the isospin triplets of $Z_c(3900)$ were established. Using the data samples of the process $e^+ e^- \to \pi^+ \pi^- \pi^0 \eta_c$ at $\sqrt{s}=4.226$ GeV, the BESIII Collaboration observed the evidence of the decay $\zc^\pm \to \rho^\pm \eta_c$~\cite{BESIII:2019rek}. Besides these hidden charm processes, the charmonium-like state $\zc$ has also been observed in the open charm process. In the $D^\ast \bar{D}$ invariant mass distributions of the process $e^{+}e^{-} \to \pi^{\pm,0} (D\bar{D}^{*}+c.c)^{\mp,0}$, the BESIII Collaboration further confirmed the existence of $\zc$~\cite{BESIII:2013qmu, BESIII:2015pqw, BESIII:2015ntl,Belle:2015yoa,BESIII:2015aym,BESIII:2017bua}.

	As a cousin of $\zc$, the charged charmonium-like state $\zcp$ was firstly observed in the $\pi^\pm h_c$ invariant mass distributions of the process $e^+ e^- \to \pi^+ \pi^- h_c$ in the year 2013~\cite{BESIII:2013ouc}, and later, its neutral partner, $\zcp^0$ was observed in the $\pi^0 h_c $ invariant mass distributions of the process $e^+ e^- \to \pi^0 \pi^0 h_c$~\cite{BESIII:2014gnk}. Similar to the case of $\zc$, the charmonium like state $\zcp$ has also been observed in the open charm process. In the year 2014, the BESIII Collaboration observed the charged $\zcp^{\pm}$ in the $(D^{*} \bar{D}^{*})^{\pm}$ invariant mass distribution of the process $e^+e^- \to  \pi^\mp (D^{*} \bar{D}^{*})^{\pm} $ at $\sqrt{s}=4.26$ GeV~\cite{BESIII:2013mhi}, later the neutral one was also observed in the open charm channel~\cite{BESIII:2015tix}.

	The experimental analyses of $\zc$ and $\zcp$ indicate that their $I^G(J^P)$ quantum numbers are $1^+(1^+)$. The isospin triplets nature shows that there are at least four constituent quarks in both $\zc$ and $\zcp$, thus, some tetraquark interpretations have been proposed ~\cite{Chen:2010ze,Voloshin:2013dpa,Chen:2015fsa,Chen:2016qju,Liu:2019zoy,Wang:2013llv,Agaev:2017tzv}. Moreover, the observed masses of $\zc$ and $\zcp$ are in the vicinities of the thresholds of $D^\ast\bar{D}$ and $D^\ast \bar{D}^\ast$, respectively, which indicate that $\zc$ and $\zcp$ could be good candidates of molecular states composed of $D^\ast \bar{D}+c.c$ and $D^\ast \bar{D}^\ast$~\cite{Wang:2013cya,Liu:2009qhy,Sun:2012zzd,Wang:2013qwa,Cui:2013xla,Wang:2013daa,Chen:2013omd,Wang:2014gwa,Aceti:2014uea,Zhao:2014gqa,Gutsche:2014zda,Aceti:2014kja,Karliner:2015ina,Chen:2015igx,Albaladejo:2015lob,Ji:2022uie}, respectively. However, the lattice estimations~\cite{Prelovsek:2014swa} and the phenomenological investigations in Ref.~\cite{He:2015mja,Du:2022jjv} indicated the interaction between $D^\ast \bar{D}$ is not strong enough to form a bound state and the observed $\zc$ could be interpreted as a $D^\ast \bar{D}$ resonance. Besides the resonance interpretations, some kinematical mechanisms have also been proposed ~\cite{Chen:2011xk,Chen:2013coa,Swanson:2014tra,Szczepaniak:2015eza,Swanson:2015bsa,BESIII:2015wge}. More details of the experimental and theoretical progresses in $\zc$ and $\zcp$ can be found in the recent reviews \cite{Guo:2017jvc,Olsen:2017bmm,Brambilla:2019esw,Ali:2017jda,Guo:2019twa,Liu:2013waa,Dong:2017gaw}.
	
	To date, the nature of the $\zc$ and $\zcp$ is still in debate. From the experimental side, searching for the evidence of $\zc$ and $\zcp$ in more processes are essential for decoding their nature. With the accumulation of the experimental data, more channels have been precisely measured in the electron-positron annihilation process by BESIII and Belle Collaborations~\cite{ParticleDataGroup:2022pth}. For example, in the year 2014, the Belle Collaboration reported their precise measurement of $e^+e^- \to \pi^+ \pi^- \psi(3686)$ via initial state radiation process, and the evidence for a charged charmonium-like structure at 4.05 GeV was observed in the $\pi^\pm \psi(3686)$ intermediate state in the $Y(4360)$ decays~\cite{Belle:2014wyt}. Later, the BESIII Collaboration reported the measurement of the same process, and in the $\pi^\pm \psi(3686)$ invariant mass distribution a charged structure was observed with a mass $m=(4032.1 \pm 2.4)$ MeV~\cite{BESIII:2017tqk}, but the experimental data can not be well described with a fixed width in different  kinematic regions.
	
	Moreover, the BESIII Collaboration measured the cross sections for $e^+ e^- \to \pi^+ \pi^-\psi(3770)$, and the $\pi^\pm \psi(3770)$ invariant distributions were also reported~\cite{BESIII:2019tdo}. In the $\pi^\pm \psi(3770)$ invariant distributions, the hints for peaks at $4.04$ and $4.13$ GeV was observed in the $\sqrt{s}=4.42$ GeV data with a rather low statistical significance. It should be noted that the mass of the peak structure at 4.04 GeV is close to the one of $\zcp$, while the peak at 4.19 GeV could be considered as the reflection structure of $\zcp$. In this case, the branching fraction of $\zcp^\pm  \to \pi^\pm \psi(3770) $ should be sizable. However, in the $\sqrt{s}=4.26$ and $4.36$ GeV data where the process $e^+ e^- \to \pi^\pm \zcp^\mp$ have been observed, no $\zcp$ structure is discovered  in the $\pi^\pm \psi(3770) $ invariant mass distributions with the present statistics. Thus, the theoretical estimations of the branching fractions of $\zcp^+ \to  \pi^+ \psi(3770)$ may shied light on the nature of the peaks in the $\pi^\pm \psi(3770) $ invariant mass distributions.
	
In addition, a large data sample of $e^+ e^- \to \pi^+ \pi^-\psi_2(3828)/\pi^0 \pi^0 \psi_2(3842)$ has also been collected by BESIII detector in recent years~\cite{BESIII:2022yga,BESIII:2022cyq}, which may also provide us with a good opportunity of searching for charged charmonium-like states in $\pi^\pm \psi_2(3823)$. Together with $\psi_3(3842)$ observed by LHCb Collaboration~\cite{LHCb:2019lnr}, the $D$-wave spin triplets charmonia have been well established. Considering the $J^P$ quantum numbers conservations and the kinematical limitation, one finds that $\zcp$ can transit to $\psi(1^3D_J), \ (J=1,2,3)$ by emitting a pion. Thus, in the present work, we estimate the branching fractions of $\zcp^+ \to \pi^+\psi(1^3D_J), \ (J=1,2,3) $ and discuss the experimental potential of observing $\zcp^\pm$ in $\pi^\pm \psi(1^3D_J)$ invariant mass distributions.

	This work is organized as follows. After introduction,  we present our estimations of the branching fractions of the process $Z_c(4020)^+ \to  \pi^+  \psi(1^3D_J)$ in Section \ref{sec:Sec2}, where the final state interactions plays the dominant role. The numerical results and related discussions are presented in Section~\ref{Sec:Num} and the last section is devoted to a short summary.

	\section{THE HIDDEN CHARM DECAYS of $Z_c(4020)$}
	\label{sec:Sec2}
	
	\begin{figure}[htb]
		\begin{tabular}{cc}
			\centering
			\includegraphics[width=4.0cm]{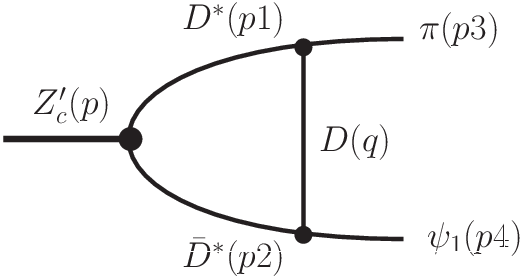}&
			\includegraphics[width=4.0cm]{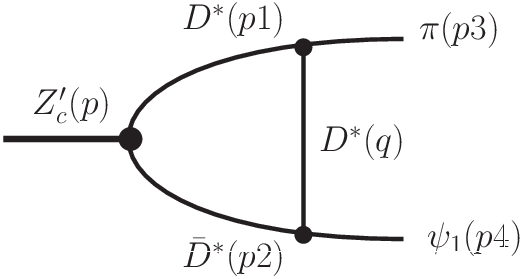}\\ \\
			$(a)$ & $(b)$ \\
		\end{tabular}
		\caption{Diagrams contributing to $Z_c(4020)\to \pi \psia$,  at the hadron level, where the momenta of the involved particles are labeled.\label{Fig:Tri1}}
	\end{figure}

	\begin{figure}[htb]
	\begin{tabular}{ccc}
		\centering
\includegraphics[width=2.8cm]{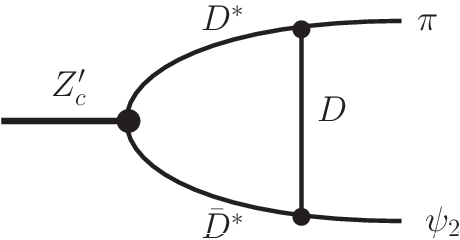}&
\includegraphics[width=2.8cm]{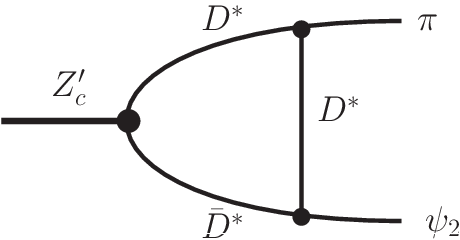}&
\includegraphics[width=2.8cm]{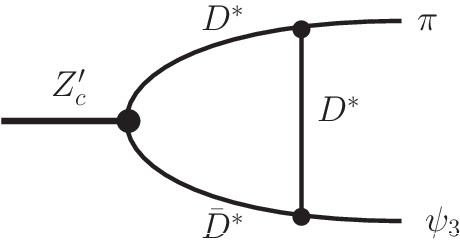}\\ \\
$(c)$ & $(d)$ & $(e)$ \\
	\end{tabular}
	\caption{Diagrams contributing to $Z_c(4020) \to \pi \psib$ (diagrams (c) and (d)) and $Z_c(4020)\to \pi \psic$ (diagrams (e)) at the hadron level.\label{Fig:Tri2}}
\end{figure}

The experimental measurement indicates that $\zcp$ dominantly decays into $D^\ast \bar{D}^\ast$\cite{BESIII:2013ouc,BESIII:2013mhi}, the $D^\ast \bar{D}^\ast$ pair can transits into a charmonium and a light meson by exchanging a proper charmed meson, such kind of meson loop mechanism can well reproduce the decay properties of $\zc$ and $\zcp$~\cite{Xiao:2018kfx,Chen:2015igx,Chen:2017abq}. Here, we further extend such decay mechanism to investigate the processes $\zcp^+ \to \pi^+ \psi(1^3D_J) ,\ (J=1,2,3)$. In Fig.~\ref{Fig:Tri1} and Fig.~\ref{Fig:Tri2}, the meson loop contributing to the relevant decay processes are presented.

	\subsection{Effective Lagrangians}
	In the present work, the meson loops are evaluated at the hadronic level, and the effective Lagrangian approach is adopted to depict the interactions between the mesons. As for the interactions between $Z_c(4020)$ (hereinafter, $Z_c^\prime$ refers to $\zcp$) and $D^\ast \bar{D}^\ast$ meson pair,  the corresponding effective Lagrangian can be expressed as,
	\begin{eqnarray}
		\mathcal{L}=i g_{Z_{c}^{\prime} D^{*} D^{*}} \varepsilon_{\mu \nu \alpha \beta} \partial^{\mu} Z_{c }^{\prime \nu} D^{\ast \alpha} \bar{D}^{\ast \beta}.\label{Eq:Lzcp}
	\end{eqnarray}
	
	As for the effective interactions relevant to the $D$ wave charmonia and charmed meson pair, they are constructed in the heavy quark limit. For the heavy-light mesons, the wave function should be independent on the flavor and spin of the heavy quarks in the heavy quark limit, and the heavy meson can be fully  characterized by the light degrees of freedom. Motivated by the nonrelativistic quark model, the light degrees of freedom can be decomposed as $\vec{s}_\ell =\vec{\ell}+\vec{s}_q$ with $\vec{\ell}$ and $\vec{s}_q$ to be the orbital angular momentum and the spin of light quark, respectively. Each value of $s_\ell$ corresponds to a degenerated doublets with total spin $S=s_\ell \pm 1/2$ and in the infinite heavy quark mass limit, the doublet degenerate in mass. For the $S$-wave charmed mesons doublets, $ \left(D,D^*\right)$  and $\left( \bar{D},\bar{D}^*\right)$  can be expressed in the matrix form, which are~\cite{Lipkin:1986av,Casalbuoni:1996pg, Kaymakcalan:1983qq,Oh:2000qr},
	\begin{eqnarray}
		H_1&=&\frac{1+v\!\!\!\slash}{2}\left[D^{\ast\mu}\gamma_\mu-D\gamma_5\right]\nonumber\\
		H_2&=&\left[\bar{D}^{\ast\mu}\gamma_\mu-\bar{D}\gamma_5\right]\frac{1+v\!\!\!\slash}{2}
	\end{eqnarray}
	respectively, and $H_{1,2}$ satisfy $\bar{H}_{1,2}=\gamma^0 H^\dagger_{1,2}\gamma^0$.

	For the heavy quarkonium, the heavy flavor symmetry is violated in the leading order, while the degeneracy is still expected under the  rotations of the two heavy quark spins, which allows us to build up multiplets for each value of the orbital angular momentum. For the $D$ wave charmonia, the matrix form of the multiplets can be expressed as~\cite{Lipkin:1986av},
	\begin{eqnarray}
		\mathcal{J}^{\mu\lambda}&=&\frac{1+v\!\!\!\slash}{2}\Bigg[\psi^{\mu\alpha\lambda}_3 \gamma_\alpha+\frac{1}{\sqrt{6}}\left(\psi^{\mu\alpha\beta\rho}v_\alpha \gamma_\beta \psi^\lambda_{2\rho}+\psi^{\lambda\alpha\beta\rho}v_\alpha \gamma_\beta \psi^\mu_{2\rho}\right)\nonumber\\
		&&+\frac{\sqrt{15}}{10}\left[ \Big(\gamma^\mu-v^\mu\Big)\psi^\lambda_1+\left(\gamma^\lambda-v^\lambda\right)\psi^\mu_1\right]\nonumber\\
		&&-\frac{1}{\sqrt{15}}\left(g^{\mu\lambda}-v^\mu v^\lambda\right)\gamma_\alpha \psi^\alpha_1+\eta^{\mu\lambda}_{c}\gamma_5\Bigg]\frac{1-v\!\!\!\slash}{2}.
	\end{eqnarray}
	
	With the above matrix expressions of the heavy light mesons and the $D$-wave heavy quarkonia, the leading order of the effective Lagrangian describing the coupling between $D$-wave charmonia and charmed meson pair can be expressed as~\cite{Li:2013xia},
	\begin{eqnarray}
		\mathcal{L}=ig_2 \mathrm{Tr}\left[\mathcal{J}^{\mu\lambda}\bar{H}_2 {\stackrel{\leftrightarrow}{\partial_\mu}}\gamma_\lambda \bar{H}_1\right]+\mathrm{H.c.}.
	\end{eqnarray}
	
	After further expanding the above Lagrangian, we can obtain the specific effective interactions relevant to the current calculation, which are,
	\begin{eqnarray}
		\mathcal{L}&=&g_{\psi_1 DD}\psi^\mu_1 \left(D \partial_\mu D^\dagger-D^\dagger\partial_\mu D\right)\nonumber\\
		&&+g_{\psi_1 DD^\ast}\varepsilon^{\mu\nu\alpha\beta}\Big[D{\stackrel{\leftrightarrow}{\partial_\mu}}D^{\ast\dagger}_\beta-D^{\ast}_\beta{\stackrel{\leftrightarrow}{\partial_\mu}}D^\dagger\Big]\partial_\nu \psi_{1\alpha}\nonumber\\
		&&+g_{\psi_1 D^\ast D^\ast}\Big[-4\left(\psi^\mu_1 D^{\ast\nu\dagger}\partial_\mu D^{\ast}_\nu-\psi^\mu_1 D^{\ast}_\nu \partial_\mu D^{\ast\nu\dagger}\right)+\psi^\mu_1 D^{\ast\nu\dagger}\partial_\nu D^{\ast}_\mu\nonumber\\
		&&-\psi^\mu_1 D^{\ast\nu}\partial_\nu D^{\ast\dagger}_\mu\Big]+ig_{\psi_2 D D^\ast}\psi^{\mu\nu}_2\left(D{\stackrel{\leftrightarrow}{\partial_\nu}}D^{\ast\dagger}_\mu-D^{\ast}_\mu{\stackrel{\leftrightarrow}{\partial_\nu}}D^\dagger\right)\nonumber\\
		&&+ig_{\psi_2 D^\ast D^\ast}\varepsilon_{\alpha\beta\mu\nu}\Big[D^{\ast\nu}{\stackrel{\leftrightarrow}{\partial^\beta}}D^{\ast\dagger}_\lambda-D^{\ast\nu\dagger}{\stackrel{\leftrightarrow}{\partial^\beta}}D^{\ast}_\lambda\Big]\partial^\mu \psi^{\alpha\lambda}_2\nonumber\\
		&&+g_{\psi_3 D^\ast D^\ast}\psi^{\mu\nu\alpha}_3 \Big[D^{\ast}_\alpha{\stackrel{\leftrightarrow}{\partial_\mu}}D^{\ast\dagger}_\nu+D^{\ast}_\nu{\stackrel{\leftrightarrow}{\partial_\mu}}D^{\ast\dagger}_\alpha\Big].\nonumber\\ 
		\label{Eq:Lag1}
	\end{eqnarray}
	
	Considering the heavy quark limit and the chiral symmetry, the effective interaction related to light pseudoscalar mesons and charm mesons are constructed as follows~\cite{Falk:1992cx,Chen:2014sra, Yan:1992gz,Cheng:1992xi,Wise:1992hn},
	\begin{eqnarray}
		\mathcal{L}_{D^{(*)} D^{(*)}\mathcal{P}}&=&-ig_{D^\ast D \mathcal{P}}\left(D^{i\dagger} \partial^\mu \mathcal{P}_{ij}D^{\ast j}_\mu-D^{\ast i\dagger}_\mu \partial^\mu \mathcal{P}_{ij}D^{j}\right)\nonumber\\
		&&+\frac{1}{2}g_{D^\ast D^\ast \mathcal{P}}\varepsilon_{\mu\nu\alpha\beta}D^{\ast \mu\dagger}_i \partial^\nu \mathcal{P}_{ij}{\stackrel{\leftrightarrow}{\partial^\alpha}}D^{\ast \beta}_j, \label{Eq:Lag2}
	\end{eqnarray}
	where $D=(D^0,D^+,D_{s}^+ )$ is the charmed meson triplets and the concrete expression of the pseudoscalar mesons in the traceless matrix is,
	\begin{eqnarray}
		\mathcal{P}&=&\left(\begin{array}{ccc}
			\sqrt{\frac{1}{2}}\pi^{0}+\sqrt{\frac{1}{6}} \eta & \pi^{+} & K^{+} \\
			\pi^{-} & -\sqrt{\frac{1}{2}} \pi^0+ \sqrt{\frac{1}{6}}\eta & K^{0} \\
			K^{-} & \bar{K}^{0} & -\sqrt{\frac{2}{3}} \eta
		\end{array}\right).\ \
	\end{eqnarray}
	
	\subsection{Decay Amplitudes}
	With the above effective Lagrangians, one can obtain the amplitudes of $Z_c(4020)^+\to\pi^+\psi(1^{3}D_J)$ corresponding to the diagrams in Fig.~\ref{Fig:Tri1} and Fig.~\ref{Fig:Tri2} are,
	\begin{eqnarray}
		\mathcal{M}_{a}&=&i^{3} \int \frac{d^{4} q}{(2 \pi)^{4}}\left[i g_{Z_{c}^{\prime}} \varepsilon_{\mu \nu \alpha \beta}(-i) p^{\mu} \epsilon_{Z_{c}^{\prime}}^{\nu}\right]\left[i g_{D^{*} D \pi} i p_{3 \theta}\right]\nonumber\\
		&&\left[-g_{\mathrm{\psi}_{1} D^{*} D} \varepsilon_{\rho \tau \sigma \xi}(-i)\left(p_{2}^{\rho}-q^{\rho}\right) i p_{4}^{\tau} \epsilon_{\mathrm{\psi}_{1}}^{\sigma}\right] \frac{-g^{\alpha \theta}+p_{1}^{\alpha} p_{1}^{\theta} / m_{1}^{2}}{p_{1}^{2}-m_{1}^{2}}\nonumber\\
		&&\times \frac{-g^{\beta \xi}+p_{2}^{\beta} p_{2}^{\xi} / m_{2}^{2}}{p_{2}^{2}-m_{2}^{2}} \frac{1}{q^{2}-m_{q}^{2}} \mathcal{F}\left(q^{2}, m_{q}^{2}\right),\nonumber\\
	\end{eqnarray}
	\begin{eqnarray}
		\mathcal{M}_{b}&=&i^{3} \int \frac{d^{4} q}{(2 \pi)^{4}}\left[ g_{Z_{c}^{\prime}} \varepsilon_{\mu \nu \alpha \beta}(-i) p^{\mu} \epsilon_{Z_{c}^{\prime}}^{\nu}\right]\nonumber\\
		&&\times \left[\frac{1}{2} g_{D^{*} D^{*} \pi} \varepsilon_{\theta \phi \kappa \lambda} i p_{3}^{\phi} i\left(p_{1}^{\kappa}+q^{\kappa}\right)\right]\nonumber\\
		&&\times \left[g _ { \psi _ { 1 } D^ { * } D^ { * } } \epsilon _ { \psi _ { 1 } } ^ { \rho }\left(-4\left(q_{\rho}-p_{2 \rho}\right) g_{\sigma \xi}+q_{\sigma} g_{\rho \xi}\right.\right.\left.\left.-p_{2 \xi} g_{\sigma \rho}\right)\right]\nonumber\\
		&&\times \frac{-g^{\alpha \theta}+p_{1}^{\alpha} p_{1}^{\theta} / m_{1}^{2}}{p_{1}^{2}-m_{1}^{2}} \frac{-g^{\beta \sigma}+p_{2}^{\beta} p_{2}^{\sigma} / m_{2}^{2}}{p_{2}^{2}-m_{2}^{2}}\nonumber\\
		&&\times \frac{-g^{\lambda \xi}+q^{\lambda} q^{\xi} / m_{q}^{2}}{q^{2}-m_{q}^{2}} \mathcal{F}\left(q^{2}, m_{q}^{2}\right) ,\nonumber\\
		%
		\mathcal{M}_{c}&=&i^{3} \int \frac{d^{4} q}{(2 \pi)^{4}}\left[i g_{Z_{c}^{\prime}} \varepsilon_{\mu \nu \alpha \beta}(-i) p^{\mu} \epsilon_{Z_{c}^{\prime}}^{\nu}\right]\left[i g_{D^{*} D \pi} i p_{3 \rho}\right]\nonumber\\
		&&\times\left[i g_{\psi_{2} D^{*} D} \epsilon_{\psi_{2}}^{\tau \xi}(-i)\left(p_{2 \xi}-q_{\xi}\right)\right] \frac{-g^{\alpha \rho}+p_{1}^{\alpha} p_{1}^{\rho} / m_{1}^{2}}{p_{1}^{2}-m_{1}^{2}}
		\nonumber\\
		&&\times \frac{-g_{\tau}^{\beta}+p_{2}^{\beta} p_{2 \tau} / m_{2}^{2}}{p_{2}^{2}-m_{2}^{2}} \frac{1}{q^{2}-m_{q}^{2}} \mathcal{F}\left(q^{2}, m_{q}^{2}\right)\nonumber\\
%
		\mathcal{M}_{d}&=&i^{3} \int \frac{d^{4} q}{(2 \pi)^{4}}\left[i g_{Z_{c}^{\prime}} \varepsilon_{\mu \nu \alpha \beta}(-i) p^{\mu} \epsilon_{Z_{c}^{\prime}}^{\nu}\right]\nonumber\\
		&&\times\left[\frac{1}{2} g_{D^{*} D^{*} \pi} \varepsilon_{\rho \tau \sigma \xi} i p_{3}^{\tau} i \times\left(p_{1}^{\sigma}+q^{\sigma}\right)\right]   \nonumber\\
		&&\times\left[i g _ { \psi _ { 2 } D^ { * } D^ { * } } \varepsilon _ { \theta \phi \kappa \lambda } \left(g_{\eta}^{\lambda} g_{\delta \omega}\left(p_{2}^{\phi}-q^{\phi}\right)-g_{\omega}^{\lambda} g_{\delta \eta}\left(q^{\phi}\right.\right.\right.
		\nonumber\\
		&&\left.\left.\left.-p_{2}^{\phi}\right)\right) p_{4}^{\kappa} \epsilon_{\psi_{2}}^{\theta \delta}\right] \frac{-g^{\alpha \rho}+p_{1}^{\alpha} p_{1}^{\rho} / m_{1}^{2}}{p_{1}^{2}-m_{1}^{2}} \frac{-g^{\beta \omega}+p_{2}^{\beta} p_{2}^{\omega} / m_{2}^{2}}{p_{2}^{2}-m_{2}^{2}} \nonumber\\
		&&\times \frac{-g^{\xi \eta}+q^{\xi} q^{\eta} / m_{q}^{2}}{q^{2}-m_{q}^{2}} \mathcal{F}\left(q^{2}, m_{q}^{2}\right),\nonumber\\
	%
		\mathcal{M}_{e}&=&i^{3} \int \frac{d^{4} q}{(2 \pi)^{4}}\left[i g_{Z_{c}^{\prime}} \varepsilon_{\mu \nu \alpha \beta}(-i) p^{\mu} \epsilon_{Z_{c}^{\prime}}^{\nu}\right]\left[\frac{1}{2} g_{D^{*} D^{*} \pi} \varepsilon_{\rho \tau \sigma \xi} i p_{3}^{\tau} i\right. \nonumber\\
		&&\left.\times\left(p_{1}^{\sigma}+q^{\sigma}\right)\right]\left[g_{\psi_{3} D^{*} D^{*}} \epsilon_{\psi_{3}}^{\theta \phi \kappa}(-i)\left(p_{2 \theta}-q_{\theta}\right)\right. \nonumber\\
		&&\left.\times\left(g_{\kappa \lambda} g_{\phi \eta}+g_{\phi \lambda} g_{k \eta}\right)\right]
		\nonumber\\
		&&\times \frac{-g^{\alpha \rho}+p_{1}^{\alpha} p_{1}^{\rho} / m_{1}^{2}}{p_{1}^{2}-m_{1}^{2}} \frac{-g^{\beta \eta}+p_{2}^{\beta} p_{2}^{\eta} / m_{2}^{2}}{p_{2}^{2}-m_{2}^{2}}\nonumber\\
		&&\times \frac{-g^{\beta \lambda}+q^{\xi} q^{\lambda} / m_{q}^{2}}{q^{2}-m_{q}^{2}} \mathcal{F}\left(q^{2}, m_{q}^{2}\right).
	\end{eqnarray}
	In the above amplitudes, a form factor ${F}(q^2,m_q^2)$ is introduced to reflect the off-shell effect of the exchange mesons and to make the amplitude convergent in the ultraviolet region. In the present work, a form factor in the monopole form is employed, which is,
	\begin{eqnarray}
		\mathcal{F}\left(q^{2},m_q^2 \right)=\left(\frac{m_q^{2}-\Lambda^{2}}{q^{2}-\Lambda^{2}}\right)^{2},
	\end{eqnarray}
	where the parameter $\Lambda$ is reparameterized as $\Lambda=m_{q}+\alpha \Lambda_{QCD}$ with $\Lambda_{QCD}=220$ MeV and $m_q$ to be the mass of the exchanged meson. The model parameter $\alpha$ should be of the order one, However, its concrete value cannot be derived by the first principle methods~\cite{Tornqvist:1993vu,Tornqvist:1993ng,Locher:1993cc,Li:1996yn}. In practice, the exact value of $\alpha$ is usually determined by comparing the experimental measurements with the theoretical estimates. However, the relevant processes discussed in the present work have not been measured, thus, we can not direct compare our estimations with the corresponding experimental measurement. Fortunately, a similar hidden charm decay process, $\zcp^+ \to \pi^+ h_c  $, has been measured, which was investigated in our previous work~\cite{Xiao:2018kfx}. In the present work, we take $\zcp^+ \to \pi^+ h_c  $ as a scale to discuss the branching fractions of $\zcp^+ \to \pi^+ \psi(1^3 D_J), \ J=(1,2,3)$. For simplify, the relevant effect Lagrangians and amplitudes for $\zcp^+ \to \pi^+ h_c  $ are not presented in this work, more details can be found in Ref.~\cite{Xiao:2018kfx}.

	In addition, the momentum of the heavy-light meson could be expressed as $p^\mu=m_Q v^\mu +k_\mu$ with $v$ to be the heavy meson velocity and $k$ is the residual momentum, thus, in the heavy quark limit, the  numerator of the propagator for the vector charmed meson can be scaled as $-g^{\mu \nu}+v^\mu v^\nu$. To further test the rationality of this approximation, we further investigate the same processes with the propagator without any approximation, which is presented in the Appendix \ref{Sec:Appendix}. Considering that the mass of the charm quark is not heavy enough, thus, in the present work, the approximation $p^\mu \simeq m_Q v^\mu$ is only adopted in the numerator of the vector meson propagators, while the momenta of the relevant charmed meson in the vertexes keep unchanged. It should be noted that the nonrelativistic treatment of the numerator of the vector charmed meson propagator consistent with the heavy quark symmetry. However, as listed in Eqs. (\ref{Eq:Lag1}) and (\ref{Eq:Lag2}), the effective Lagrangians with Levi-Civita tensor function have an additional partial differential operator resulted from $v^\mu$, which makes the effective Lagrangian break the heavy quark symmetry in a certain extend. In particular, this leads to different asymptotic behavior of the amplitudes in the ultraviolet region, and the model parameter dependence for different channels may be a bit different.

With the above amplitudes corresponding the meson loop diagrams in Fig.~\ref{Fig:Tri1} and Fig.~\ref{Fig:Tri2}, we can obtain the amplitudes for $\zcp^+ \to\pi^+ \psi(1^{3}D_J)$, which are,
	\begin{eqnarray}
		\mathcal{M}_{Z_{c}^{\prime}\to \psi(1^3D_1) \pi} &=& 2(\mathcal{M}_a +\mathcal{M}_b),\nonumber\\
		\mathcal{M}_{Z_{c}^{\prime}\to \psi(1^3D_2) \pi}&=& 2(\mathcal{M}_c +\mathcal{M}_d),\nonumber\\
		\mathcal{M}_{Z_{c}^{\prime}\to \psi(1^3D_3) \pi}&=& 2\mathcal{M}_e,
	\end{eqnarray}
	where the factor 2 comes from the charge symmetry.
	Then the decay widths of $\zcp^+ \to\pi^+ \psi(1^{3}D_J)$ could be estimated by
	\begin{eqnarray}
		\Gamma(Z_c^\prime \to \psi_J \pi)=\frac{1}{3} \frac{1}{8 \pi} \frac{|\vec{p}|}{m_{Z_c^\prime}^{2}} \overline{\left|\mathcal{M}_{Z_c^\prime \to \psi(1^{3}D_J) \pi}\right|^{2}},
	\end{eqnarray}
	where $\vec{p}$ denotes the momentum of daughter particles in the initial rest frame.

	\section{Numerical Results and discussion}
	\label{Sec:Num}
	\subsection{Coupling constants}
	So far, $\zcp$ has only been observed in the $D^\ast \bar{D}^\ast$ and $\pi h_c$ channels. At $\sqrt{s}=4.26$ GeV, the Born cross sections for $e^+ e^- \to \pi^\pm (D^\ast \bar{D}^\ast)^\mp$ are measured to be $(137\pm 9\pm 15)$ pb and the fraction from the quasi-two body cascade decay process $e^+e^- \to \pi^\pm \zcp^\mp \to \pi^\pm (D^\ast \bar{D}^\ast)^\mp$  was measured to be $0.65 \pm 0.09\pm 0.06$~\cite{BESIII:2013mhi}, while the cross sections for the quasi-two body process $e^+e^-\to \pi^\pm \zcp^\mp \to \pi^+ \pi^- h_c$ was measured to be $(7.4\pm 1.7 \pm 2.1)$ pb~\cite{BESIII:2013ouc}. Then, the ratio of the widths of $\zcp \to D^\ast \bar{D}^\ast$ and $\zcp \to h_c\pi$ is estimated to be,
	\begin{eqnarray}
		\frac{\Gamma[\zcp\to D^\ast \bar{D}^\ast]}{\Gamma[\zcp \to h_c \pi]}=12.0 \pm 3.68\pm 3.48. \label{Eq:Ratioexp}
	\end{eqnarray}
	With the central values of the estimated ratio and the width of $\zcp$, one can find the central value of $\Gamma[\zcp \to D^\ast \bar{D}^\ast]$ is about 12 MeV by assuming that $\zcp$ dominantly decays into $D^\ast \bar{D}^\ast$ and $\pi h_c$. By comparing the width of $\zcp \to D^\ast \bar{D}^\ast$ estimated by the effective Lagrangian in Eq.~(\ref{Eq:Lzcp}) and the one deduced by the experimental measurements, one can obtain the coupling constant  $g_{Z_{c}^{\prime}}$=1.13~\cite{ParticleDataGroup:2022pth}.

	In the heavy quark limit, the coupling constants of the $D-$wave charmonia and the charmed meson pair satisfy,
	\begin{eqnarray}
		g_{\psi_1 DD}&=&-2g_2 \frac{\sqrt{15}}{3}\sqrt{m_{\psi_1}m_D m_D},\nonumber\\
		g_{\psi_1 DD^\ast}&=&g_2 \frac{\sqrt{15}}{3}\sqrt{m_{D}m_{D^\ast}/m_{\psi_1}},\nonumber\\
		g_{\psi_1 D^\ast D^\ast}&=&-g_2 \frac{\sqrt{15}}{15}\sqrt{m_{\psi_1}m_{D^\ast} m_{D^\ast}},\nonumber\\
		g_{\psi_2 D D^\ast}&=&2g_2 \sqrt{\frac{3}{2}}\sqrt{m_{\psi_2}m_D m_{D^\ast}},\nonumber\\
		g_{\psi_2 D^\ast D^\ast}&=&2g_2 \sqrt{\frac{1}{6}}\sqrt{m_{D^\ast}m_{D^\ast}/m_{\psi_2}},\nonumber\\
		g_{\psi_3 D^\ast D^\ast}&=&2g_2 \sqrt{m_{\psi_3}m_{D^\ast} m_{D^\ast}},
		\end{eqnarray}
	where $g_2$ is a gauge coupling and its value can be estimated by the measured width of $\psi(3770)\to D\bar{D}$, which is $g_2=1.39~\mathrm{GeV}^{-3/2}$.
	
	Considering the heavy quark limit and chiral symmetry, we can obtain that the coupling constants related to the pion meson satisfy,
	\begin{eqnarray}
		g_{D^{\ast}D\pi}&=&\frac{2g}{f_\pi}\sqrt{m_{D^\ast}m_{D}},\nonumber\\
		g_{D^{\ast}D^\ast \pi}&=&\frac{2g}{f_\pi}.
	\end{eqnarray}
	with $f_\pi=132 \mathrm{MeV}$ to be the decay constant of pion meson, $g=0.55$, which was determined by using the partial width of $D^\ast\rightarrow D\pi$~\cite{ParticleDataGroup:2022pth}.

\subsection{Decay width of $\zcp^+ \to \pi^+ h_c $}

With the above preparations, All the relevant parameters have been fixed expect for the model parameter $\alpha$, which is introduced by the form factor in the amplitudes. In the present work, we try to determine the parameter range by reproducing the width of $\zcp^+ \to \pi^+ h_c$. With the assumption that $\zcp$ dominantly decay into $D^\ast \bar{D}^\ast$ and $\pi h_c $ and the measured ratio in Eq.~(\ref{Eq:Ratioexp}), we find that the partial width of $\zcp^+\to \pi^+  h_c$  is $(0.72\sim 1.63)$ MeV. By comparing our estimations with the data extracted from experimental measurements, we find that the estimated results overlap with the experimental data in the $\alpha$ range of 1.34-1.87. With this parameter range, we can further discuss the decay processes $\zcp^+ \to \pi^+ \psi(1^3D_J),~J=(1,2,3) $.

\begin{figure}[t              ]
		\centering
		\includegraphics[width=8.4cm]{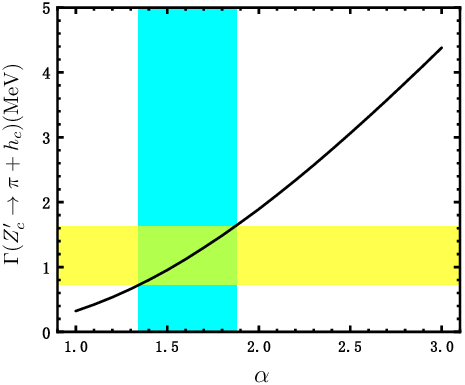}
		\caption{(Color online). The width of $\zcp^+ \to \pi^+ h_c$ depending on the model parameter $\alpha$. The yellow horizontal band is the partial width extracted from the experimental measurements, while the cyan vertical band is the $\alpha$ range determined by the partial width of $Z_{c}^\prime\to h_c+\pi$.\label{Fig:Bhc3}}
	\end{figure}

\subsection{Branching fractions of $\zcp^+\to \pi^+ \psi(1^3D_J),~J=(1,2,3) $}
	
\begin{figure}
		\centering
	\includegraphics[width=8.4cm]{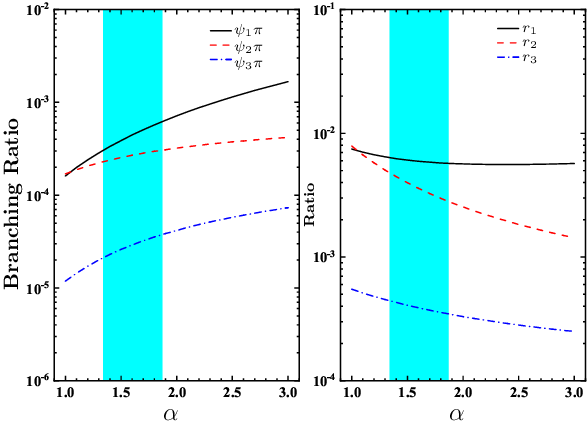}
		\caption{(Color online). The branching fractions of $Z_c(4020)^+ \to \pi^+ \psi(1^{3}D_J)$ (left panel) and the ratio R of Eq.~(\ref{con:r}) (right panel) depending on the model parameter $\alpha$.}\label{Fig:Br}
	\end{figure}

In the present work, we take $\zcp^+ \to \pi^+ h_c  $ as a scale, and discuss the decay width of $\zcp^+ \to \pi^+ \psi(1^3 D_J)),~J=(1,2,3) $ in the same model parameter range. With the centered value of the width of $\zcp^+$, we can further estimate the branching fractions of $\zcp^+ \to \pi^+\psi(1^3D_J),~J=(1,2,3)$ depending on the model parameter $\alpha$.	The branching fractions of $\zcp^+ \to \pi^+ \psi(1^3D_J),~J=(1,2,3)$ depending on the model parameter $\alpha$ are presented in Fig~\ref{Fig:Br}.
From the figure one can find that the branching fractions for $\zcp^+ \to \pi^+ \psi(1^3D_J) ,~J=(1,2,3)$ increase with the model parameter $\alpha$. With the parameter range determined by $\zcp^+ \to \pi^+ h_c $, we find that the branching fractions of $Z_c(4020)^+ \to \pi^+  \psi(1^{3}D_J),~J=(1,2,3)$ are,
	\begin{eqnarray}
		B[Z_c(4020)^+\to\psi(3770)\pi^+] &=& (3.03-6.17)\times10^{-4}\nonumber\\
		B[Z_c(4020)^+\to\psi_2(3823)\pi^+] &=& (2.28-3.04)\times10^{-4}\nonumber\\
		B[Z_c(4020)^+\to\psi_3(3842)\pi^+]&=&(2.10-3.37)\times10^{-5},\ \ \ \
	\end{eqnarray}	
expected from the power counting, which are much smaller than the one of $\zcp^+ \to \pi^+ h_c $.

As shown in Fig.~\ref{Fig:Br}, the branching fractions for $\zcp^+ \to \pi^+ \psi(1^3D_J) ,~J=(1,2,3)$ and $\zcp^+ \to \pi^+ h_c$ are predicted with large uncertainties, but all these branching fractions increase with the model parameter $\alpha$, thus, one can expect the ratios of these branching fractions should be weakly dependent on the model parameter $\alpha$. Thus, we can further discuss the experimental potential of observing $\zcp^+ \to \pi^+ \psi(1^3D_J),~J=(1,2,3)$ by the ratios of the branching fractions of $\zcp^+ \to \pi^+ \psi(1^3D_J),~J=(1,2,3)$ and $\zcp^+ \to \pi^+ h_c$, which are estimated to be,
	\begin{eqnarray}
		r_{1}&=&\frac{B\left[Z_c(4020)^+\to\pi^+ \psia \right]}{B\left[Z_c(4020)^+\to \pi^+ h_c\right]} =(4.92-5.46)\times10^{-3},\nonumber\\
		r_{2}&=&\frac{B\left[Z_c(4020)^+\to \pi^+ \psib \right]}{B\left[Z_c(4020)^+\to \pi^+ h_c\right]}=(2.42-4.12)\times10^{-3},\nonumber\\
		r_{3}&=&\frac{B\left[Z_c(4020)^+\to \pi^+ \psic \right]}{B\left[Z_c(4020)^+\to \pi^+ h_c\right]}=(2.98-3.80)\times10^{-4},\nonumber\\  \label{con:r}
	\end{eqnarray}
in the considered parameter range. From the above estimation, one can find that the ratios are estimated to be of the order of $10^{-4}$ or $10^{-3}$.

To further test the dependences of the form factor and the approximation of the propagators, we further estimated the branching fractions of $\zcp^+ \to \pi^+ \psi(1^3D_J)  ,~J=(1,2,3)$ with a full propagator for the vector charmed mesons and a form factor in the form as shown in Eq.~(\ref{Eq:App-FFs}). As indicated in Ref.~\cite{Xiao:2018kfx}, the parameter range determined by $\zcp^+ \to \pi^+ h_c $ is $2.07 \sim 2.75$, with this parameter range, the branching fractions of $\zcp^+ \to \pi^+ \psi(1^3D_J),~J=(1,2,3)$ are estimated, which are listed in Eq. (\ref{Eq:App-BF}). Our estimations indicate that branching fractions of $\zcp^+ \to \pi^+ \psi_2(3823)$ and $\zcp^+ \to \pi^+ \psi_3(3842)$ estimated in two kinds of form factor are very similar, while the one of $\zcp^+ \to \pi^+ \psi(3770)$ are a bit different. Additionally, the ratios of $\zcp \to \pi^+ \psi(1^3D_J) ,~J=(1,2,3)$ and $\zcp^+ \to \pi^+ h_c$ are estimated to be of the order of $10^{-4}$, $10^{-3}$ and $10^{-2}$ for $\psi_3(3842)$, $\psi_2(3823)$ and $\psi(3770)$, respectively.

\subsection{Power counting analysis of the meson loops}
Besides the above estimations, we can also perform a power counting analysis to discuss the meson loops contributions \cite{Guo:2010ak, Guo:2010zk, Guo:2017jvc}. In the following, we take $Z_c(4020)^+\to\pi^+ \psia$ and $Z_c(4020)^+\to \pi^+ h_c$ as examples. In the diagrams of Fig.~\ref{Fig:Tri1}, both the $D^{(*)}\bar{D}^{(*)}\pi$ and $\psia D^{*}\bar{D}^{(*)}$ vertices have a P-wave coupling, while the $Z_c$ couples to the $D^{(*)}\bar{D}^{(*)}$ in a S-wave. According to the power counting, the amplitudes from the diagrams in Fig.~\ref{Fig:Tri1} scale as
\begin{equation}
\mathcal{A}_A \sim N_A\, \frac{v_A^5 \vec{q}^2_A}{(v_A^2)^3 m^2_D} =  N_A \frac{\vec{q}^2_A}{v_A m^2_D}, \label{Eq:A1}
\end{equation}
where $N_A$ collects all the constant factors, e.g., the coupling constants, the loop geometrical factor and the normalization factors, and a factor of $1/m^2_D$
with $m_D$ being the charm meson mass is introduced to balance the dimension of $\vec{q}^2$.

 As for $Z_c(4020)^+\to \pi^+ h_c$, the diagrams could be obtained by replacing the $\psia$ with $h_c$ in Fig.~\ref{Fig:Tri1}. In this case, both the $Z_c D^* \bar{D}^*$ and $h_c D^{*}\bar{D}^{(*)}$ vertices have a $S$-wave coupling, while the $\pi$ couples to the $D^{(*)}\bar{D}^{(*)}$ in $P$-wave. As a result, the corresponding amplitudes scale as,
\begin{equation}
\mathcal{A}_B \sim N_B\, \frac{v_B^5 \vec{q}_B}{(v_B^2)^3 m_D} =  N_B \frac{\vec{q}_B}{v_B m_D}, \label{Eq:A2}
\end{equation}
where $N_B$ collects all the constant factors, and a factor of $1/m_D$ is introduced to balance the dimension of $\vec{q}$.

With the amplitude scalings presented in Eqs.~(\ref{Eq:A1}) and (\ref{Eq:A2}), we roughly estimate the ratio of the meson loop contributions for the processes $Z_c(4020)^+\to\pi^+ \psia$ and $Z_c(4020)^+\to \pi^+ h_c$, which is
\begin{eqnarray}
\frac{\mathcal{A}_A}{\mathcal{A}_B}\sim \frac{\vec{q}^2_A}{ m_D \vec{q}_B} = 0.05,
\label{Eq:AmpR}
\end{eqnarray}
assuming $N_A\sim N_B$ which is reasonable as long as all the couplings take natural values. The branching fractions ratio for $Z_c(4020)^+\to\psi(3770)\pi^+$ and $Z_c(4020)^+\to h_c\pi^+$ is proportional to the $(\mathcal{A}_A/\mathcal{A}_B)^2$, which is comparable with the estimated $r_1$ in Eq. (\ref{con:r}).

In conclusion, our estimations indicate that the branching fractions of $\zcp^+ \to \pi^+ \psi(1^3D_J) ,~J=(1,2,3)$ are rather small, and searching $\zcp^+$ in the $ \pi^+ \psi(1^3D_J),~J=(1,2,3)$ invariant mass distributions are rather difficult. Thus, the peak structures around 4.04 and 4.13 GeV in the $\pi^\pm \psi(3770)$ invariant mass distributions of $e^+ e^- \to \pi^+ \pi^- \psi(3770)$ process at $\sqrt{s}=4.42$ GeV should not come from the contributions of $\zcp$. Further precise experimental measurements of the $e^+ e^- \to \pi^+ \pi^- \psi(3770)$ process are needed to decode the nature of the peak structures at 4.04 and 4.13 GeV in the $ \pi^\pm \psi(3770)$ invariant mass distributions.

	\ \ \

	\section{SUMMARY}
	About one decade ago, the BESIII Collaboration reported two near thresholds charmonium like states, $\zc$ and $\zcp$. However, their internal structure have not been well decoded until now. Searching for more decay and production process will help us to reveal the nature of these charmonium-like states. With the accumulation of the experimental data, the BESIII Collaboration have observed the cross sections for $e^+ e^- \to \pi^+ \pi^- \psia$ and  $e^+ e^- \to \pi^+ \pi^- \psi_2(3823)/\pi^0 \pi^0 \psi_2(3823)$. The hints for the peaks at 4.04 and 4.13 GeV has been observed in the $\pi^\pm \psia $ invariant mass distributions. Inspired by the observations of BESIII Collaboration, we investigate charmed meson loop contributions to the decay process of $\zcp^+ \to  \pi^+ \psi(1^3D_J),~J=(1,2,3)$ by using an effective Lagrangian approach in the present work.

Our estimations indicate that the branching fractions of $\zcp^+ \to \pi^+ \psi(1^3D_J) ,~J=(1,2,3)$ are of the order of $10^{-4}$ for $\psi(3770)$ and $\psi_2(3823)$, and of $10^{-5}$ for $\psi_3(3843)$, respectively. Thus, we can conclude that searching $\zcp^\pm $ in the $\pi^\pm \psi(1^3D_J) ,~J=(1,2,3)$ invariant mass distributions of the $e^+ e^- \to \pi^+ \pi^- \psi(1^3D_J),~J=(1,2,3)$ process should be impossible. Hence the peak structures at 4.04 GeV and 4.19 GeV in the $\pi^\pm \psia $ invariant mass distributions should not come from the contribution of $\zcp^\pm$, and further precise experimental measurements of the $e^+ e^- \to \pi^+ \pi^- \psi(3770)$ process are needed to decode the nature of these two peak structures.

Before the end of this work, it is worth mentioning that in the present estimation we connect the final $\pi \psi(1^3D_J),~J=(1,2,3) $ with $\zcp$ via charmed meson loops, where $\zcp$ has strong coupling with $D^\ast \bar{D}^\ast$. Such kinds of meson loop mechanism should play the dominant role in the hidden charm decays of $\zcp$ in the molecular scenario, while in the tetraquark picture, $\zcp$ could directly decay into hidden charm states via the rearrangement of the four constituent quarks, which should be the major contributions to the hidden charm decays of $\zcp$. Thus, the success of meson loop mechanism in the hidden charm decays of $\zcp$ could shed light on the molecular nature of $\zcp$ in a certain extend.

\section*{ACKNOWLEDGMENTS}
	This work is supported by the National Natural Science Foundation of China under the Grant No. 12175037 and 12335001.
	
\appendix	
\section{The estimations with full form propagator for vector charmed mesons}
	\label{Sec:Appendix}

In Ref. \cite{Xiao:2018kfx}, a full form of the propagator for the vector charmed-meson is adopted. To avoid the divergence in the ultraviolet region, a  form factor in the form,
\begin{eqnarray}
	\mathcal{F}\left(q^{2},m_q^2 \right)=\left(\frac{m_q^{2}-\Lambda^{2}}{q^{2}-\Lambda^{2}}\right)^{3} ,\label{Eq:App-FFs}
\end{eqnarray}
is introduced in the amplitudes.

\begin{figure}[htb]
	\centering
	\includegraphics[width=8.4cm]{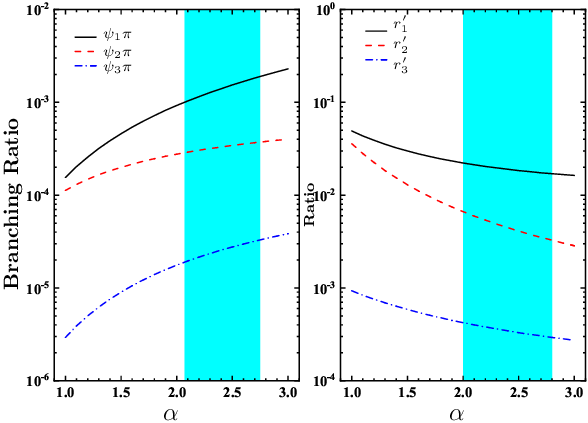}
	\caption{(Color online). The branching fractions of $Z_c(4020)^+ \to\pi^+ \psi(1^{3}D_J)$ (left panel) and the ratio defined in  Eq.~(\ref{con:rp}) (right panel)  depending on the model parameter $\alpha$.}\label{Fig:Brt}
\end{figure}

In Ref. ~\cite{Xiao:2018kfx}, the branching fraction of $\zcp^+ \to \pi^+ h_c$ has been estimated with the meson loop mechanism and the partial width could be well reproduce in the parameter range $2.07<\alpha<2.75$.  With the above form factor, the branching fractions of $\zcp^+ \to \pi^+ \psi(1^3D_J) $ are estimated, which are presented in Fig.~\ref{Fig:Brt}. In the same model parameter range, one gets the branching fractions of $\zcp^+ \to \psi(1^3D_J)\pi^+$ are	\begin{eqnarray}
	B^{\prime}[Z_c(4020)^+ \to\pi^+ \psi(1^{3}D_1)] &=& (1.00-1.90)\times10^{-3}\nonumber\\
	B^{\prime}[Z_c(4020)^+ \to \pi^+\psi(1^{3}D_2)] &=& (2.86-3.73)\times10^{-4}\nonumber\\
	B^{\prime}[Z_c(4020)^+ \to \pi^+ \psi(1^{3}D_3)]&=&(1.89-3.30)\times10^{-5}. \label{Eq:App-BF}\ \ \ \  \nonumber\\
\end{eqnarray}	

The ratios of the branching fractions for $\zcp^+ \to \pi^+ \psi(1^3D_J) $ and $\zcp \to \pi^+ h_c $ are estimated to be,
\begin{eqnarray}
	r^{\prime}_{1}&=&\frac{B\left[Z_c(4020)\to\psia \pi\right]}{B\left[Z_c(4020)\to h_c\pi\right]} =(1.51-1.81)\times10^{-2},\nonumber\\
	r^{\prime}_{2}&=&\frac{B\left[Z_c(4020)\to\psib \pi\right]}{B\left[Z_c(4020)\to h_c\pi\right]}=(2.97-5.17)\times10^{-3},\nonumber\\
	r^{\prime}_{3}&=&\frac{B\left[Z_c(4020)\to\psic \pi\right]}{B\left[Z_c(4020)\to h_c\pi\right]}=(2.62-3.42)\times10^{-4}.\nonumber\\  \label{con:rp}
\end{eqnarray}

\end{document}